\documentclass[12pt,preprint]{aastex}   

\shorttitle{Structure of shock front in SN~1006}
\shortauthors{Bamba et al.}

\begin{document}

\title{Fine Structures of Shock of SN~1006 with the {\it Chandra} Observation}

\author{
Aya Bamba, Ryo Yamazaki, Masaru Ueno, and Katsuji Koyama
}
\affil{
Department of Physics, Graduate School of Science, Kyoto University, 
Sakyo-ku, Kyoto 606-8502, Japan
}
\email{bamba@cr.scphys.kyoto-u.ac.jp,
yamazaki@tap.scphys.kyoto-u.ac.jp,
masaru@cr.scphys.kyoto-u.ac.jp,
koyama@cr.scphys.kyoto-u.ac.jp}

\begin{abstract}

The north east shell of SN~1006 is the most probable acceleration site
of high energy electrons (up to $\sim$ 100~TeV)
with the Fermi acceleration mechanism at the shock front.
We resolved non-thermal filaments from thermal emission in the shell
with the excellent spatial resolution of {\it Chandra}.
The thermal component is extended widely over about $\sim 100$~arcsec
(about 1~pc at 1.8~kpc distance) in width,
consistent with the shock width derived from the Sedov solution.
The spectrum is fitted with a thin thermal plasma of $kT$~=~0.24~keV
in non-equilibrium ionization (NEI), typical for a young SNR.
The non-thermal filaments are likely thin sheets with the scale widths of
$\sim 4$~arcsec (0.04~pc) and $\sim 20$~arcsec (0.2~pc)
at upstream and downstream, respectively.   
The spectra of the filaments are fitted with a power-law function of index
2.1--2.3, with no significant variation from position to position.
In a standard diffusive shock acceleration (DSA) model,
the extremely small scale length in upstream requires
the magnetic field nearly perpendicular to the shock normal.
The injection efficiency ($\eta$) 
from thermal to non-thermal electrons around the shock front 
is estimated to be $\sim 1\times 10^{-3}$
under the assumption that the magnetic field in upstream is 10~$\mu$G.
In the filaments,
the energy densities of the magnetic field and non-thermal electrons
are similar to each other,
and both are slightly smaller than that of thermal electrons.
in the same order for each other.
These results suggest that 
the acceleration occur in more compact region
with larger efficiency than previous studies.

\end{abstract}


\section{Introduction}

Since the discovery of cosmic rays \citep{hess},
the origin and acceleration mechanism up to $10^{15.5}$~eV
(the ``knee'' energy) have been long-standing problems.
A breakthrough came from the X-ray studies of  SN~1006;
\citet{koyama1995} discovered synchrotron X-rays from the shells of this SNR,
indicating the existence of extremely high energy electrons
up to the knee energy produced by 
the first order Fermi acceleration.
Further, \citet{tanimori1998} confirmed the presence of high energy electrons
with the detection of the TeV $\gamma$-rays,
which are cosmic microwave photons up-scattered by high energy electrons 
(the inverse Compton process) in the north east shell
(the NE shell) of SN~1006.
The combined analysis of the synchrotron X-rays
and inverse Compton TeV $\gamma$-rays nicely
reproduces  the observed flux and spectra,
and predicts a rather weak magnetic field of 4--6~$\mu$G
\citep{tanimori1998,tanimori2001}.

Since these discoveries,
detection of synchrotron X-rays and/or TeV $\gamma$-rays,
from other shell-like SNRs has been accumulating:
G347.3$-$0.5 \citep{koyama1997,slane1999,muraishi,enomoto},
RCW~86 \citep{bamba,borkowski2001b}, and G266.6$-$1.2 \citep{slane2001}.
These discoveries provide good evidence for the cosmic ray acceleration
at the shocked shell of SNRs.  
The mechanism of the cosmic ray acceleration has also been studied
for a long time
and the most plausible process is a diffusive shock acceleration (DSA)
\citep{bell, blandford1978, drury1983,blandford1987,jones,malkov}.

Apart from the globally successful picture of DSA,
detailed but important processes,
such as the injection, magnetic field configuration,
and the reflection of accelerated particles,
have not yet been well understood.
The spatial distribution of accelerated particles
responsible for the non-thermal X-rays,
may provide key information on these unclear subjects.
Previous observations, however, are limited in spatial resolution
for a detailed study on the structure of shock acceleration process
and injection efficiency.
Although many observations and theoretical models are made for SN~1006,
these problems are still open issue
\citep{reynolds1998,aharonian,vink,ellison,dyer,allen,berezhko}.

In this paper, we report on the first results
of the spectral and spatial studies on
the thermal and non-thermal shock structure in the NE shell of SN~1006
with  {\it Chandra} (\S~\ref{analyses}).
In \S~\ref{discuss1} and \S~\ref{discuss2},
we discuss the spectral analyses and determine the scale widths of the 
structures for thermal and non-thermal electrons
on the base of a simple DSA with shock parallel magnetic field.
We also derive the injection efficiency ($\eta$) 
of non-thermal electrons from the thermal plasma near the shock front.
Based on these results, we discuss possible implications on the DSA  
process in the NE shell of SN~1006. 
In this paper, we assume the distance of SN~1006 to be 1.8~kpc \citep{green}.
  
\section{Observation}

We used the {\it Chandra} archival data of the ACIS
on the NE shell of SN~1006 (Observation ID = 00732)
observed on July 10--11, 2000 with the targeted position at
(RA, DEC) = ($15^{\rm h}03^{\rm m}51\fs6$, $-41^{\rm d}51^{\rm m}18\fs8$).
The satellite and instrument are described
by \citet{weisskopf} and \citet{garmire}, respectively.   
CCD chips I2, I3, S1, S2, S3, and S4 were used with the pointing center on S3.
Data acquisition from the ACIS was made in the Timed-Exposure Faint mode
with the readout time of 3.24~s.
The data reductions and analyses were made using the {\it Chandra} 
Interactive Analysis of Observations (CIAO) software version 2.2.1.
Using the Level 2 processed events provided by the pipeline
processing at the {\it Chandra} X-ray Center,
we selected  {\it ASCA} grades 0, 2, 3, 4, and 6, as the X-ray events.
High energy electrons due to charged particles and 
hot and flickering pixels were removed.
The effective exposure was $\sim 68$~ks for the observation.
In this paper, we concentrated on the data of S3 (BI chip) 
because this chip has the best efficiency in soft X-rays  
required for the spectral analyses
and its on-axis position provides the best point-spread function required
for the spatial analysis.

\section{Analyses and Results}
\label{analyses}

\subsection{Overall Image}
\label{image}

\begin{figure}[hbtp]
\epsscale{0.9}
\plotone{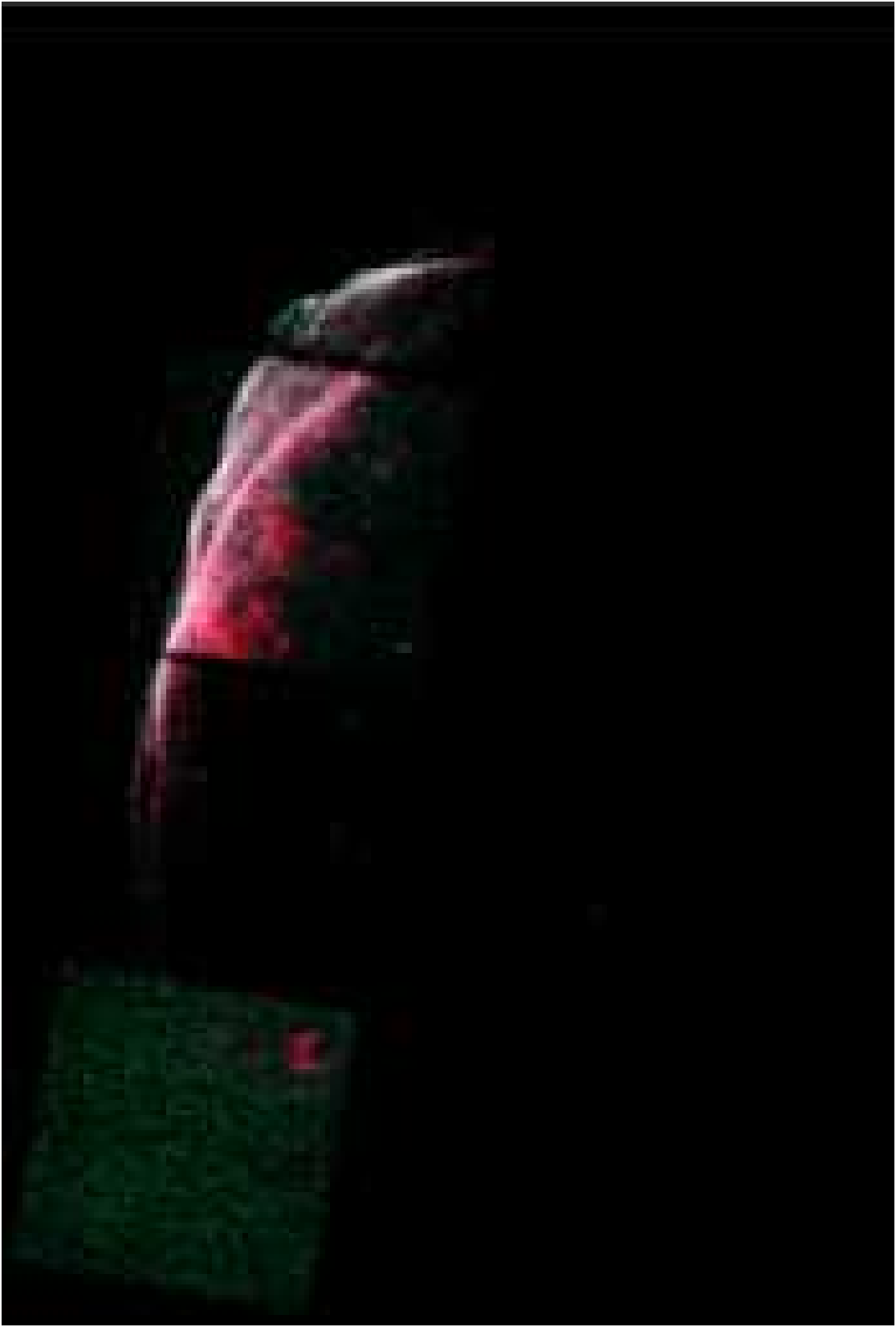}
\caption{The true-color images of SN~1006 NE shell
binned with 1~arcsec scale.
Red and blue are 0.5-2.0~keV and 2.0--10.0~keV, respectively,
both in logarithmic scale.
\label{images}}
\end{figure}

Figure~\ref{images} shows the true-color image from I2, I3, and S1--S4 
for the NE shell of SN~1006.
The image is contrasted
in the 0.5--2.0~keV band (hereinafter, the soft1 band; red) and 
in the 2.0--10.0~keV band (hereinafter, the hard band; blue)
and binned to a resolution of 1~arcsec.
The fine spatial resolution of {\it Chandra} unveils  
extremely narrow filaments in the hard band.   
They are running from north to south along the outer edge of the NE shell, 
parallel to the shock fronts observed
by H$\alpha$ emission line \citep{winkler}.
These filaments resemble the sheet-like structure of the shock
simulated by \citet{hester}.
The soft1 band image, on the other hand,  
has a larger scale width
similar to the {\it ROSAT} HRI image \citep{winkler}.
Many clumpy sub-structures are also seen in this energy band.

\subsection{Inner Shell Region}
\label{sec_thermal}

\begin{figure}[hbtp]
\epsscale{0.9}
\plotone{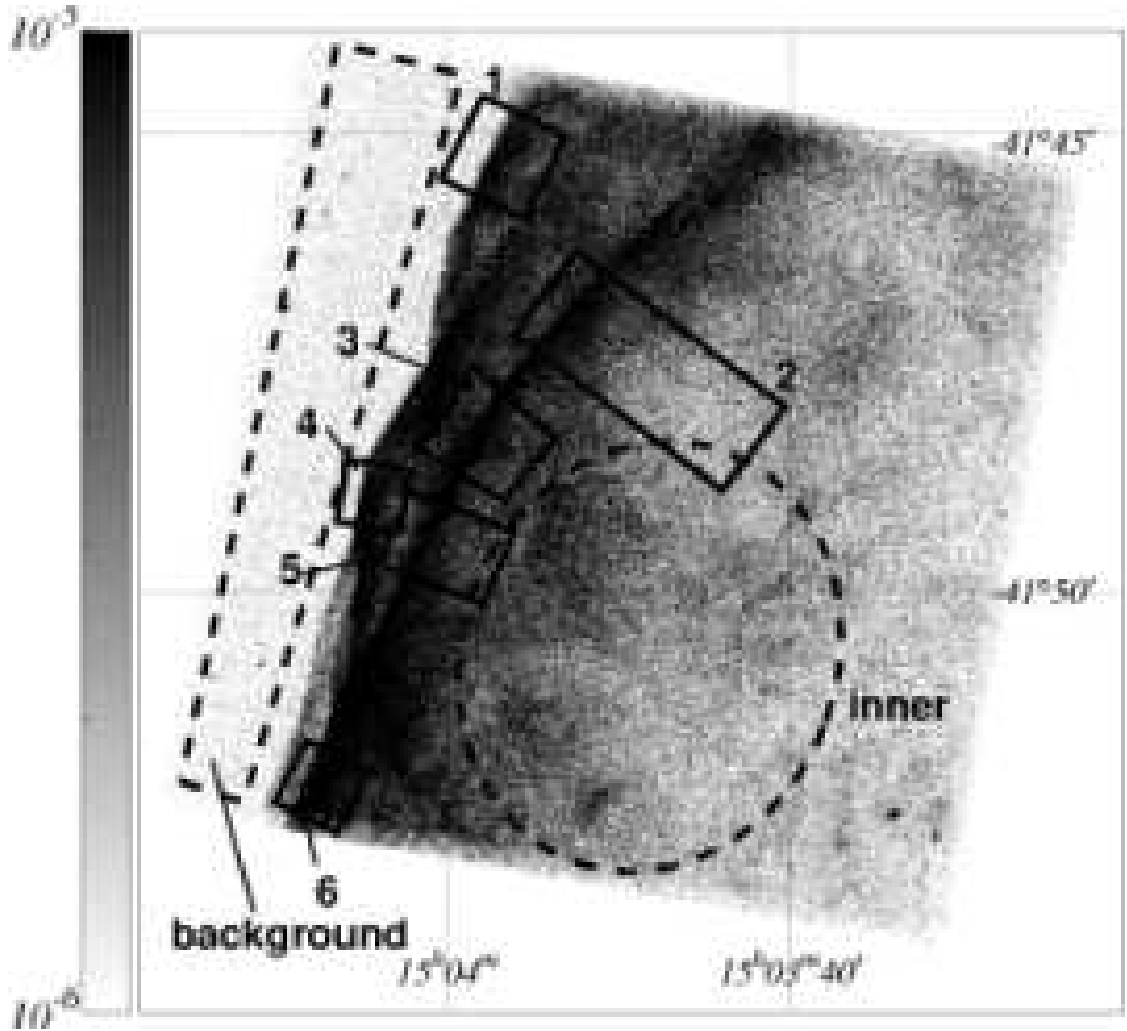}
\caption{The close-up view of the 0.5--10.0~keV band image
of S3 chip with J2000 coordinates,
binned with 1~arcsec scale.
The gray  scale (the left bar) is given
logarithmically  ranging  from $1\times 10^{-6}$ to
$1\times 10^{-5}$~cnts~s$^{-1}$~arcsec$^{-2}$ .
The inner and background regions for the spectral analyses
and the filament regions for the spatial analyses (No.1--6) are shown 
with dashed and solid lines, respectively.
\label{regions}}
\end{figure}

To resolve the thermal and non-thermal components,
we made a spectrum from a bright clump found in the soft1
band image, which is located in the inner part of the 
NE shell (``Inner region'' with the dashed ellipse in Figure ~\ref{regions}).
The background region was selected from a region out of the SNR,
as is shown in Figure~\ref{regions} with the dashed lines.

\begin{figure}[hbtp]
\epsscale{0.5}
\plotone{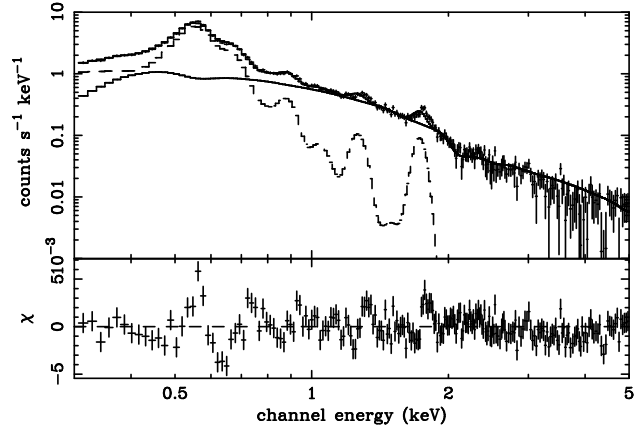}
\caption{The upper panel:
The background-subtracted spectrum of the inner region (crosses).
Dashed line and solid liens are the best-fit thin thermal and power-law
models, respectively.
The lower panel:
The data residuals from the best-fit two-components model.
\label{spectrum}}
\end{figure}

The background-subtracted spectrum shown in Figure~\ref{spectrum}
has many emission lines.
We hence determined the peak energies of the 5 brightest lines
with a phenomenological model, a power-law continuum plus Gaussian lines. 
The most strong line structures are the peak
at 0.55~keV and the hump at 0.67~keV. 
These energies are nearly equal to the  K$\alpha$ and Lyman $\alpha$ 
lines of He- and H-like oxygen,
hence are attributable to highly ionized oxygen.
Likewise, the other clear peaks at 0.87, 1.31, and 1.76~keV are
most likely He-like K$\alpha$ of Ne, Mg and Si, respectively.  
However, in detail all the observed line energies
are systematically smaller than those of the relevant atomic data.
These apparent energy shifts have been usually observed in a young SNR plasma
in non-equilibrium ionization (NEI). 
The ``energy shift'' in this case is due to the different line ratio
of many sub-levels and/or different ionization states.
The oxygen Lyman $\alpha$ is isolated  from the other lines of different 
ionization states, hence the NEI effect gives no energy shift.
Still we see apparent down-shift of the observed line energy from that of the 
laboratory data.
He-like K$\alpha$ lines are complex of many fine structures with the 
split-energy of at most $\sim$~25~eV (for He-like silicon). 
Although the energy shift of 
He-like K$\alpha$ lines due to NEI should be smaller than this split-energy, 
the observed energy shifts
are  systematically larger than the split-energy.
We therefore regard that the apparent energy shifts are due mainly
to energy calibration errors,
hence fine-tuned the energy gain to reduce by 3.8\%,
the average shift of the 5 brightest lines.
We then fitted the spectrum with a thin thermal plasma model
in NEI calculated by \citet{borkowski2001a}.
The abundances of C, N, O, Ne, Mg, Si, S, and Fe in the plasma were treated
to be free parameters, 
whereas those of the other elements were fixed to the solar values 
\citep{anders}.
The absorption column was calculated using the cross sections by
\citet{morrison} with the solar abundances.
Since this NEI model exhibited systematic data excess
at high energy above 2~keV,
we added a power-law component and the fit improved dramatically.
Figure~\ref{spectrum} and Table~\ref{thermal_para} show
the best-fit models
(dashed and solid lines for thermal and power-law components)
and parameters, respectively.

Instead of the phenomenological power-law model, we applied {\it srcut}
in the XSPEC package as a more physical model. 
Details of the {\it srcut} model fitting are given in \S~\ref{srcut}.   
The best-fit $\nu_{\rm rolloff}$ is 9.2 (8.6--10.3) $\times 10^{16}$~Hz,
with better reduced $\chi^2$ of 389.0/215
than that of the power-law model of 447.9/215
(see Table~\ref{thermal_para}).

As for the thermal components, we also tried the fitting 
with a plane shock model (XSPEC model $vpshock$) plus either
a power-law or {\it srcut} and found no essential difference
from the case of an NEI model.

Although these simple models globally follow the data very well,
all are rejected in the statistical point of view,
leaving wavy residuals near the line structure
as shown in Figure~\ref{spectrum} (lower panel).
This may be caused by improper response function
in energy scale and/or in energy resolution.
We assumed that the photons are uniformly distributed
in flux and in temperature through the whole source region.
This simple assumption may also partly responsible for the above 
systematic error,
because in reality the source region is apparently clumpy
(see Figure~\ref{images})
and may have different temperatures, abundances, and/or ionization time scales.
Since our principal aim of this paper is
to examine the spatial structures and the spectra of 
non-thermal component,
we do not examine in further detail for the thermal model.
In the following analyses and discussion,
we use the physical parameters cited in Table~\ref{thermal_para}
as a good approximation.

The spectrum of the inner region clump  is softer 
than any other regions in the NE shell,
which indicates that the contribution of the thermal component is the largest.
Nevertheless the thermal photons  are only 0.02\% of the non-thermal ones
if we limit the energy band to 2.0--10.0~keV (the hard band);
the non-thermal photons in the hard band are
8.9$\times 10^{-2}$~cnts~s$^{-1}$,
while thermal photons are 2.0$\times 10^{-5}$~cnts~s$^{-1}$.
Therefore, in the following spatial analyses,
we regard that all the photons in the hard band
are non-thermal origin.

As for the spatial analysis of the thermal emission,
we use the limited band of 0.4--0.8~keV (hereafter; the soft2 band)
to optimize the signal-to-noise ratio,
in which K-shell lines from He-like oxygen (0.57 keV) contribute
a most fraction of the X-ray emission (see Figure~\ref{spectrum}).
Even in this thermal-optimized band, however,
the count rates of the thermal and non-thermal emissions are comparable:
thermal photons are 8.1$\times 10^{-1}$~cnts~s$^{-1}$,
while those of non-thermal are 5.4$\times 10^{-1}$~cnts~s$^{-1}$.  

\subsection{The Filaments}

The outer edge of the NE shell is outlined by several thin X-ray filaments.
For the study of these filaments, we selected 6 rectangle regions
in Figure~\ref{images},
in which the filaments are straight and free from other structures
like another filament and/or clumps. 
These regions (solid boxes) are shown in Figure~\ref{regions}
with the designations of No.1--6 from north to south.
Since the SNR shell is moving (expanding) from the right to the left,
we call the right and left side 
as downstream and upstream following the terminology of the shock 
phenomena. 

Figure~\ref{profiles} shows the intensity profile
in the hard (2.0--10.0~keV: upper panel)
and soft2 (0.4--0.8~keV: lower panel) bands for each filament
with the spatial resolution of 0.5~arcsec,
where the horizontal axis ($x$-coordinate) runs from the east to west
(upstream to downstream) along the line normal to the filaments. 
We see very fast decay in the downstream side
and even faster rise in the upstream side.

To estimate the scale width, we define a simple empirical model
as a function of position ($x$) for the profiles;

\begin{equation}
f(x) = \left\{
	\begin{array}{rlr}
	A\exp |\frac{x_0-x}{w_{\rm u}}| & {\rm in\ upstream} \\
	A\exp |\frac{x_0-x}{w_{\rm d}}| & {\rm in\ downstream},
	\end{array}
\right.\label{model}
\end{equation}
where $A$ and $x_0$ are the flux and position at the filament peak,
respectively.
The scale widths are given by $w_{\rm u}$ and $w_{\rm d}$
for  upstream and downstream, respectively
(hereafter, ``u'' and ``d'' represent
upstream and downstream, respectively).
Since the scale width of the filaments is larger than
the spatial resolution of {\it Chandra}
($\sim 0.5$~arcsec = 1~bin in Figure~\ref{profiles}),
we ignore the effect of the point-spread function.

\subsubsection{Non-Thermal Structure}	

As already noted in \S~\ref{sec_thermal},
the hard band (2.0--10.0~keV) flux is nearly pure non-thermal origin
(see also the next paragraph). 
We therefore used the hard band profiles for the study of
the non-thermal X-ray structures.
The hard band profiles were fitted with a function $f^{\rm h}(x)+C^{\rm h}$
(hereafter, ``h'' represents the hard X-ray band),  
where $C^{\rm h}$ is the background constant,
which includes the cosmic and Galactic X-ray background and non X-ray events.
The fittings were statistically accepted for all the filament profiles.
The best-fit models and parameters are shown in
Figure~\ref{profiles} with the solid lines
and in Table~\ref{pro_para}, respectively.

We then made the spectra of the filaments within the
scale widths:
($x_0^{\rm h}-w_{\rm u}^{\rm h} \leq x \leq x_0^{\rm h}+w_{\rm d}^{\rm h}$)
in Figure~\ref{profiles}.
The background spectra were made from the off-filament downstream regions.
All the background-subtracted spectra are featureless (no line structure)
and extend to high energy side,
which were fitted with an absorbed power-law model
with the best-fit parameters given in Table~\ref{spec_para}.
We thus confirm that
the hard X-ray profiles represent those of non-thermal X-rays.   

To increase statistics, we summed all the data of 6 filaments
(the combined-filament). 
The best-fit power-law model parameters
for the spectrum of the combined-filament are listed in Table~\ref{spec_para}.
The spectrum was also fitted with a {\it srcut} model (see \S~\ref{srcut}).
The best-fit $\nu_{\rm rolloff}$ and the other parameters
are also listed in Table~\ref{spec_para}.
We note that the {\it srcut} model give slightly better $\chi^2$/d.o.f.
than that of a phenomenological power-law model (see Table~\ref{spec_para}).

We further spatially divided the combined-filament and analyzed each spectrum.
We however found no significant difference
between the downstream and upstream, nor within the downstream;
the photon index was nearly constant along the $x$-axis
in the combined-filament.

\subsubsection{Thermal Structure}	

\begin{figure}[hbtp]
\epsscale{0.3}
\plotone{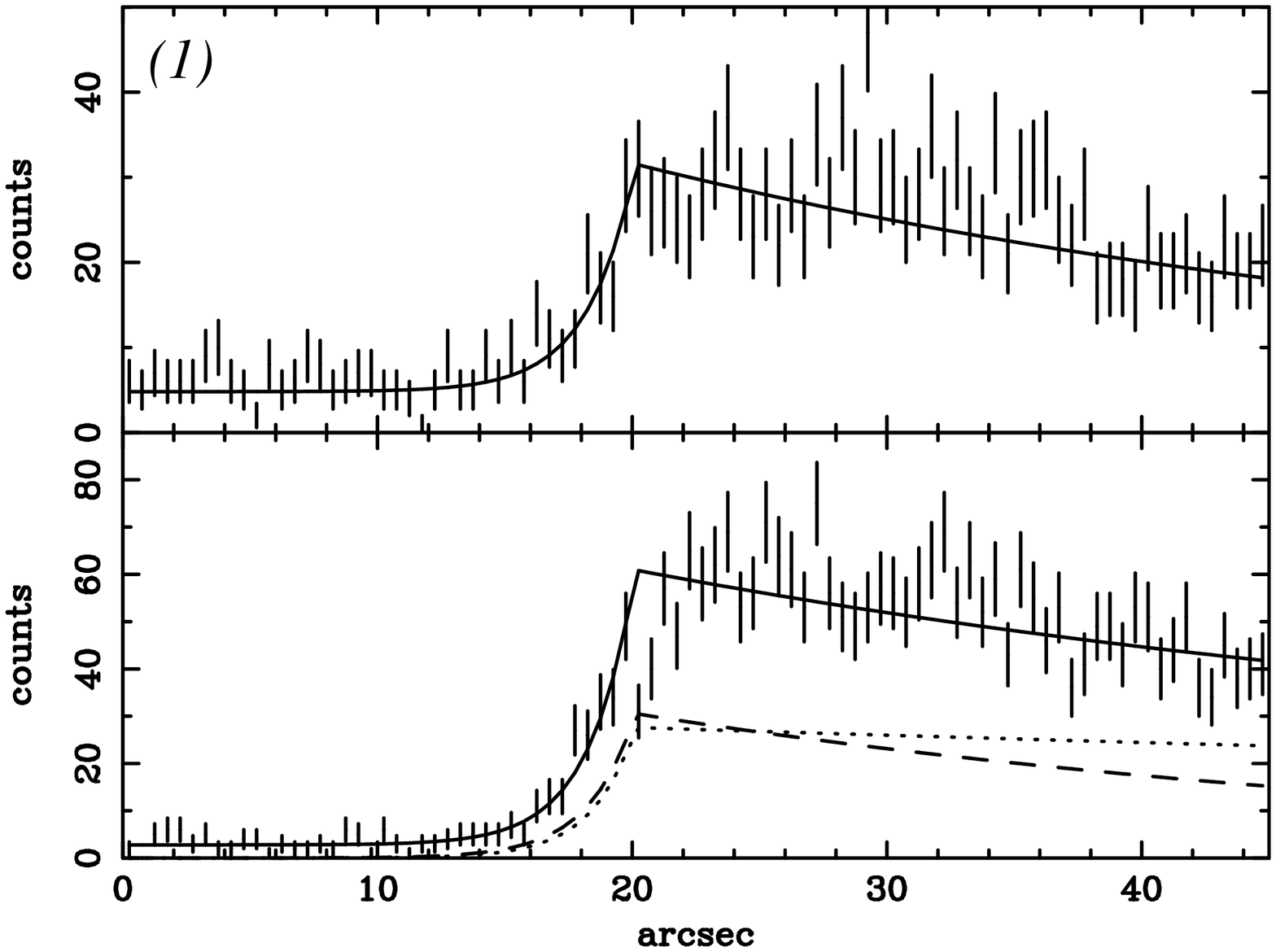}
\plotone{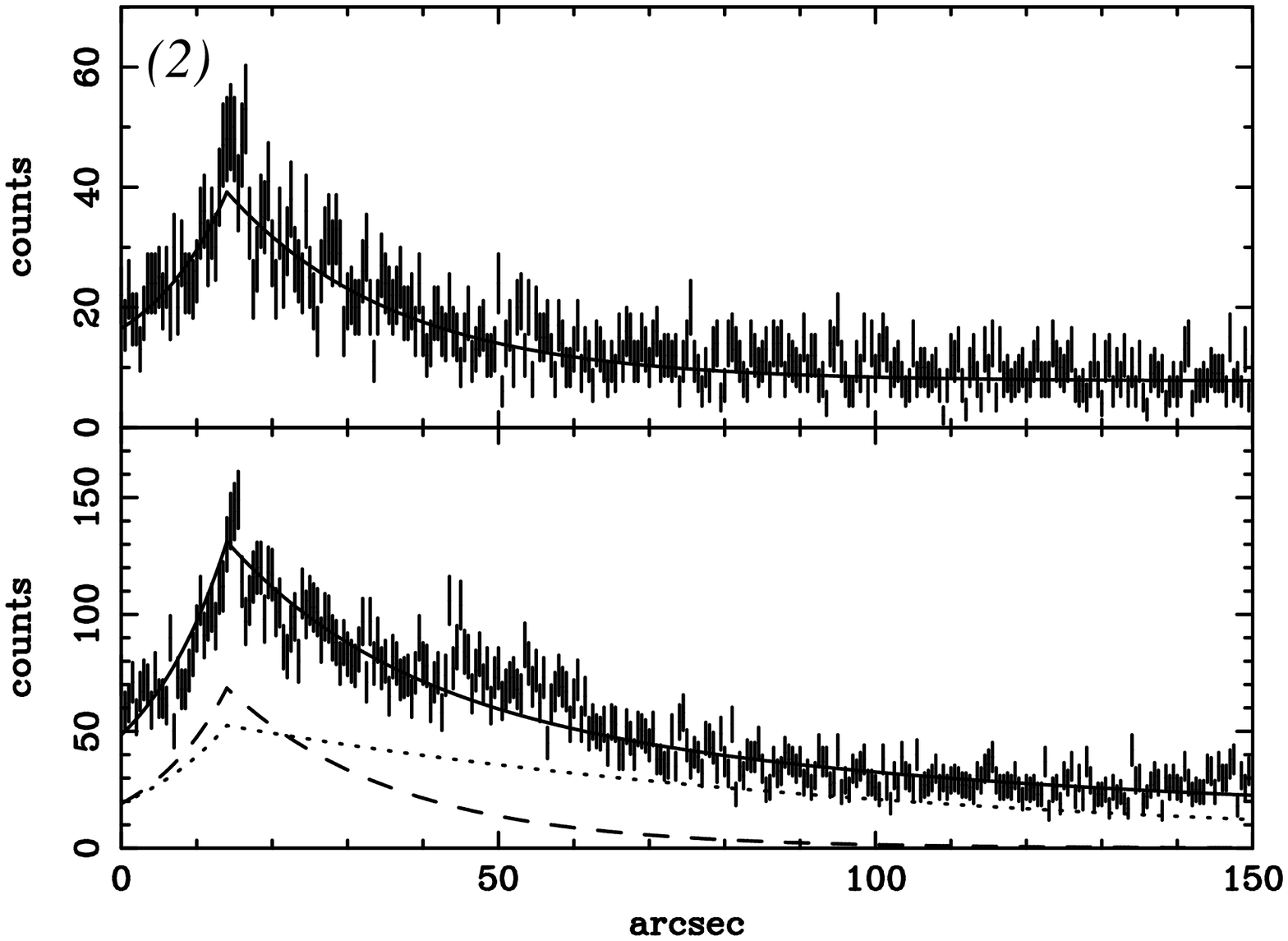}
\plotone{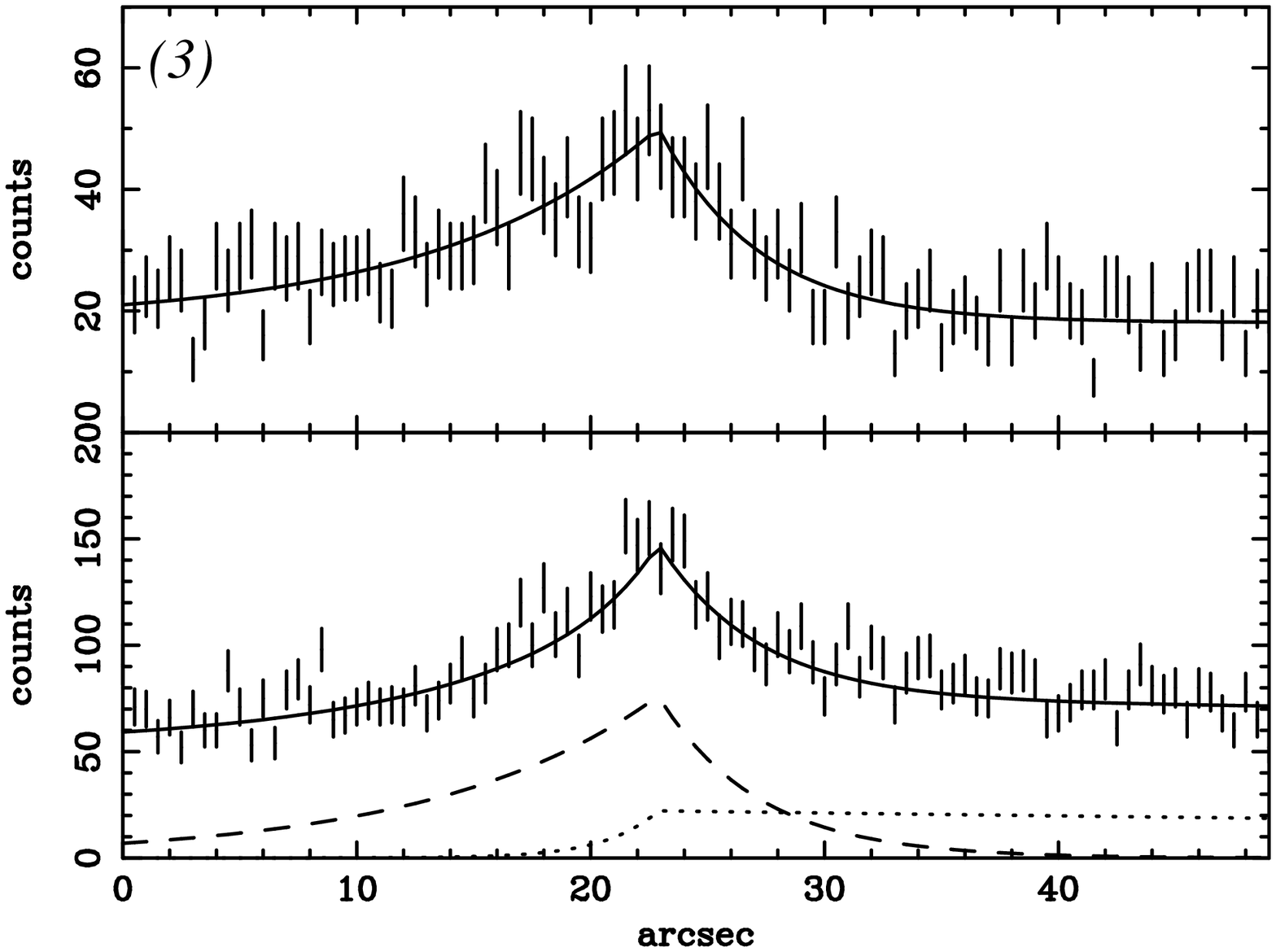}
\plotone{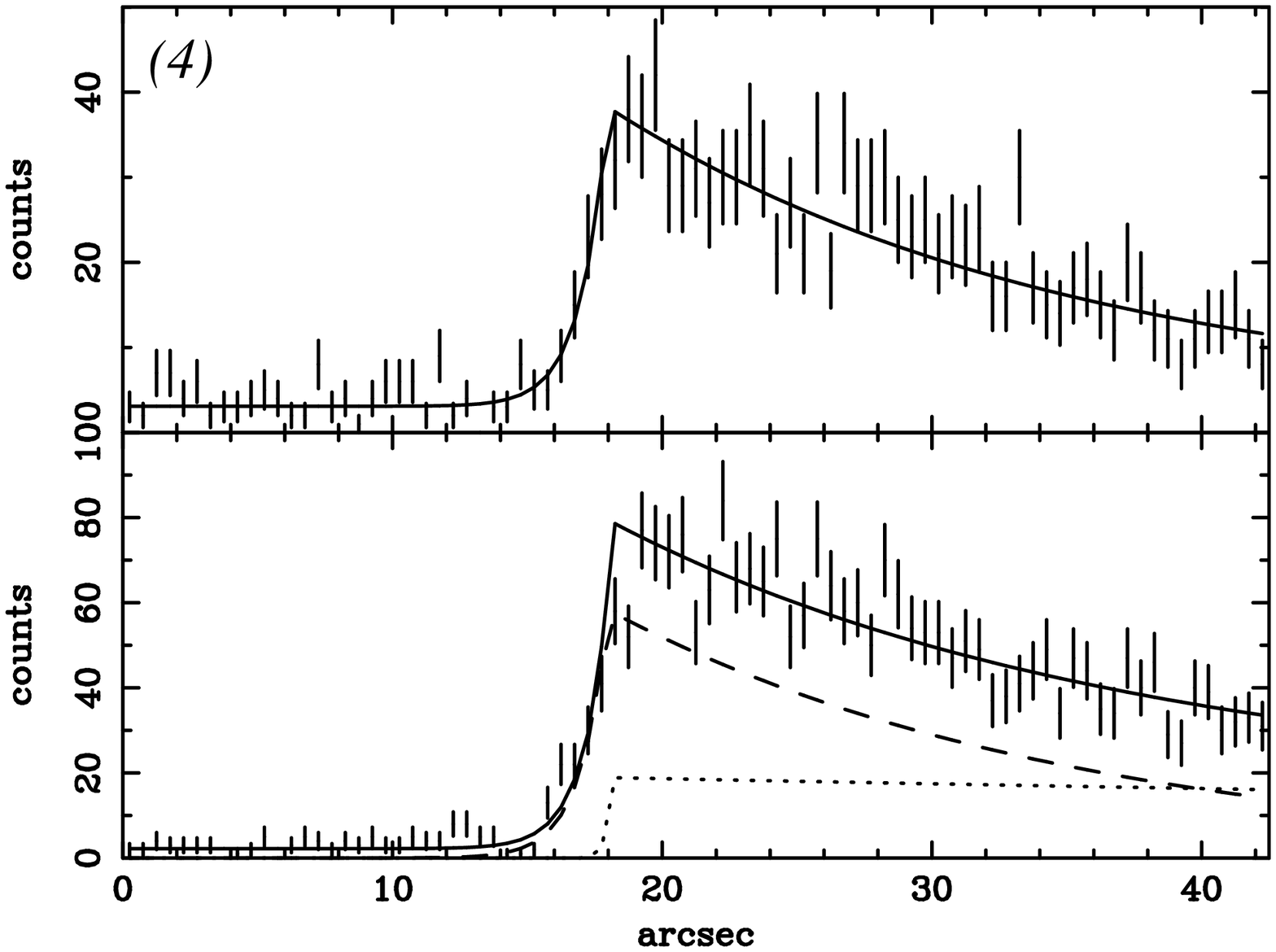}
\plotone{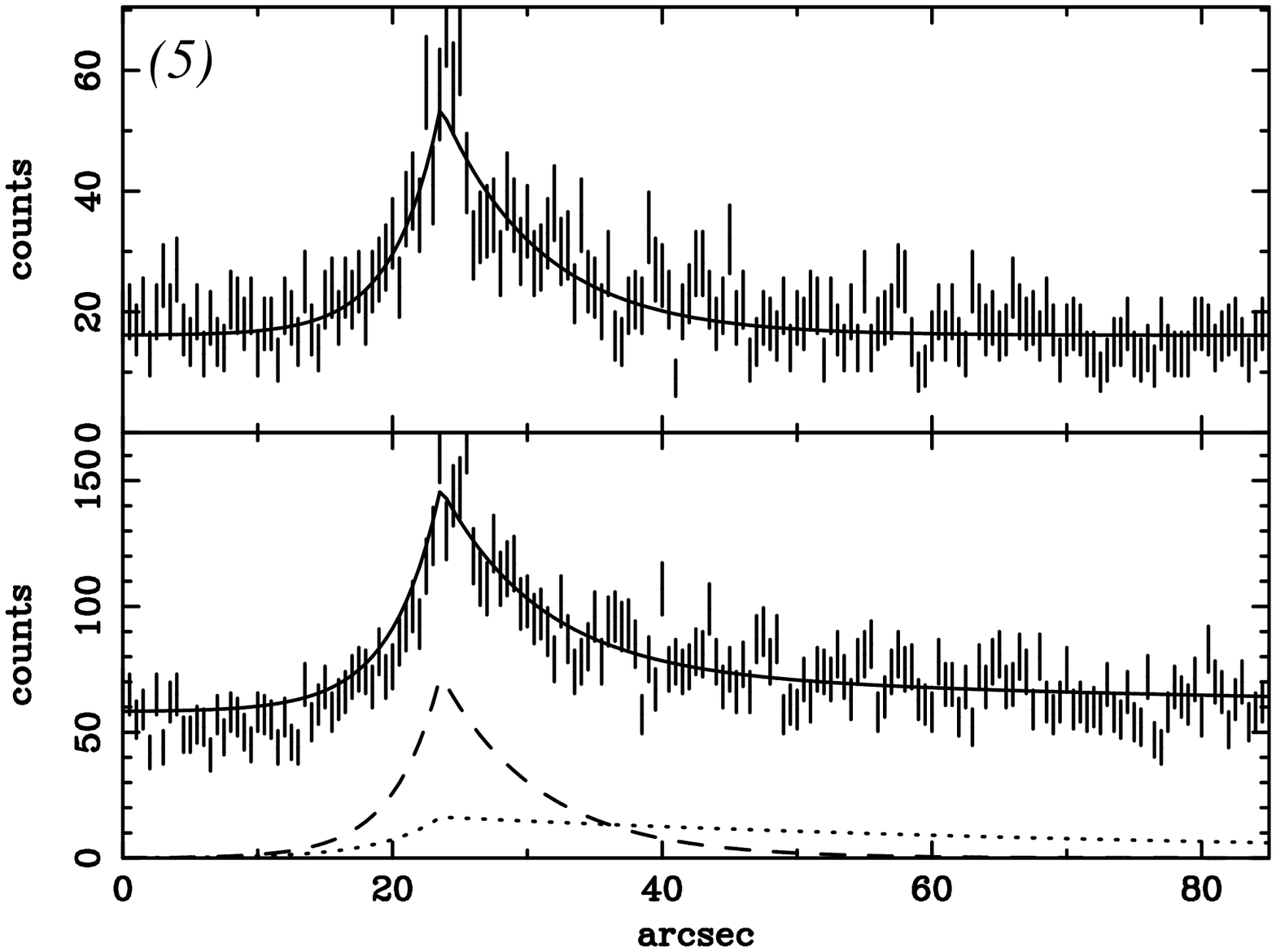}
\plotone{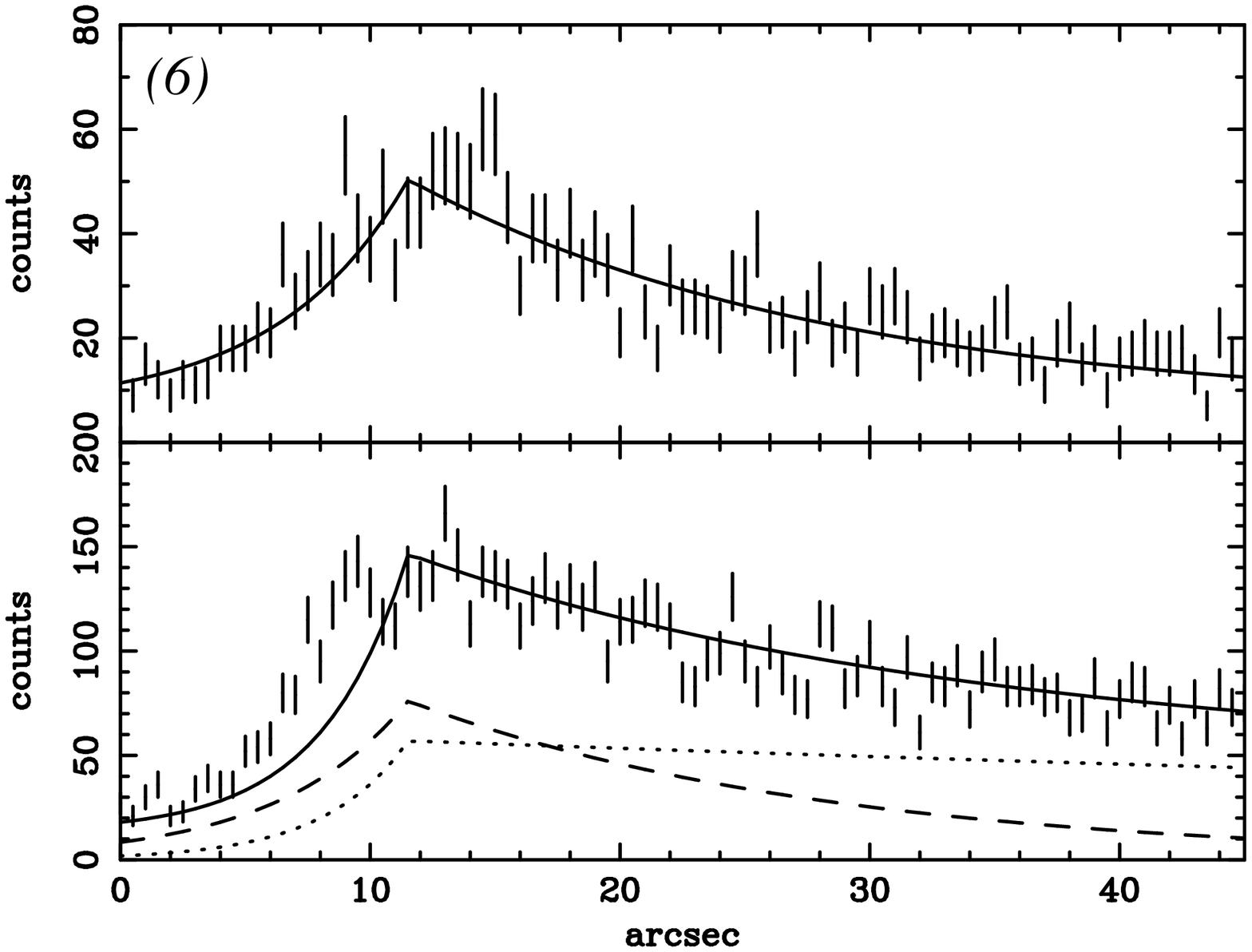}
\caption{The profiles of the filaments in SN~1006 NE shell.
Upper panels show the profiles in the hard (2.0--10.0~keV) band, 
whereas the lower panels in the soft2 (0.4--0.8~keV) band
with the best-fit models (solid lines).
The dashed lines in the lower panels represent
non-thermal photons
extrapolated from the hard band flux 
of the power-law (see the upper panels). The dotted lines
are the thermal component after subtracting the non-thermal
contamination (dashed lines).
Upstream is to the left and downstream is to the right.
\label{profiles}}
\end{figure}

To examine the structure of the thermal components,
we used the soft2 band profiles (the lower panels of Figure~\ref{profiles}).
Contamination from the non-thermal photons
would be very large even in this optimized band (see \S~\ref{sec_thermal}).
Therefore we calculated the flux ratio of the non-thermal photons
between in the hard and  soft2 bands
using the spectral parameters given in Table~\ref{spec_para}.
From the flux ratio and the best-fit hard band profiles,
we estimated the non-thermal contaminations as are
given with the dashed lines in Figure~\ref{profiles}.

After subtracting these non-thermal contaminations,
we fitted the soft2 band profiles with a model of $f^{\rm s}(x)+C^{\rm s}$,
where $C^{\rm s}$ is the background constant in the same sense as $C^{\rm h}$.
Note that although we
use ``s'' as the abbreviation of the soft2 band, it actually represents
thermal X-rays.   

From Figure~\ref{profiles},
we see that most of the photons at the filament peak is non-thermal origin,
hence the  statistics becomes poor to determine the position of the 
thermal peak ($x_0^{\rm s}$) independently.
We thus fixed $x_0^{\rm s}$ to the best fit peak
in the hard hand $x_0^{\rm h}$.
Also we set the upper bound of the fitting parameter
$w_{\rm d}^{\rm s}$ to be 900~arcsec,
the same as the radius of SN~1006 \citep{green}.
The best-fit models and parameters for the thermal components
($f^{\rm s}(x)$) are shown 
with dotted lines in the lower panels of Figure~\ref{profiles}
and in Table~\ref{pro_para}, respectively.

\subsubsection{Non-Thermal versus Thermal} 

\begin{figure}[hbtp]
\epsscale{1.0}
\plottwo{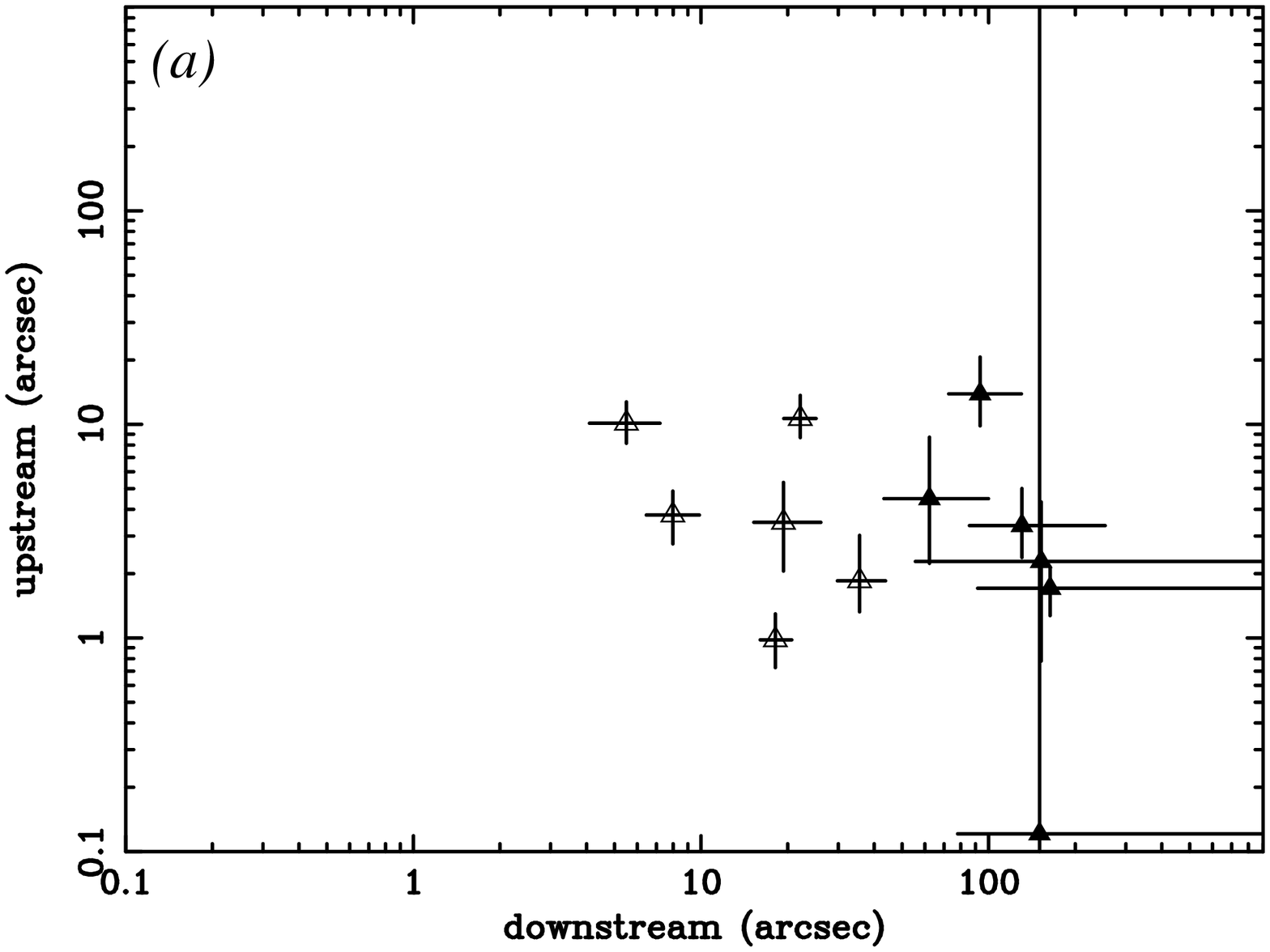}{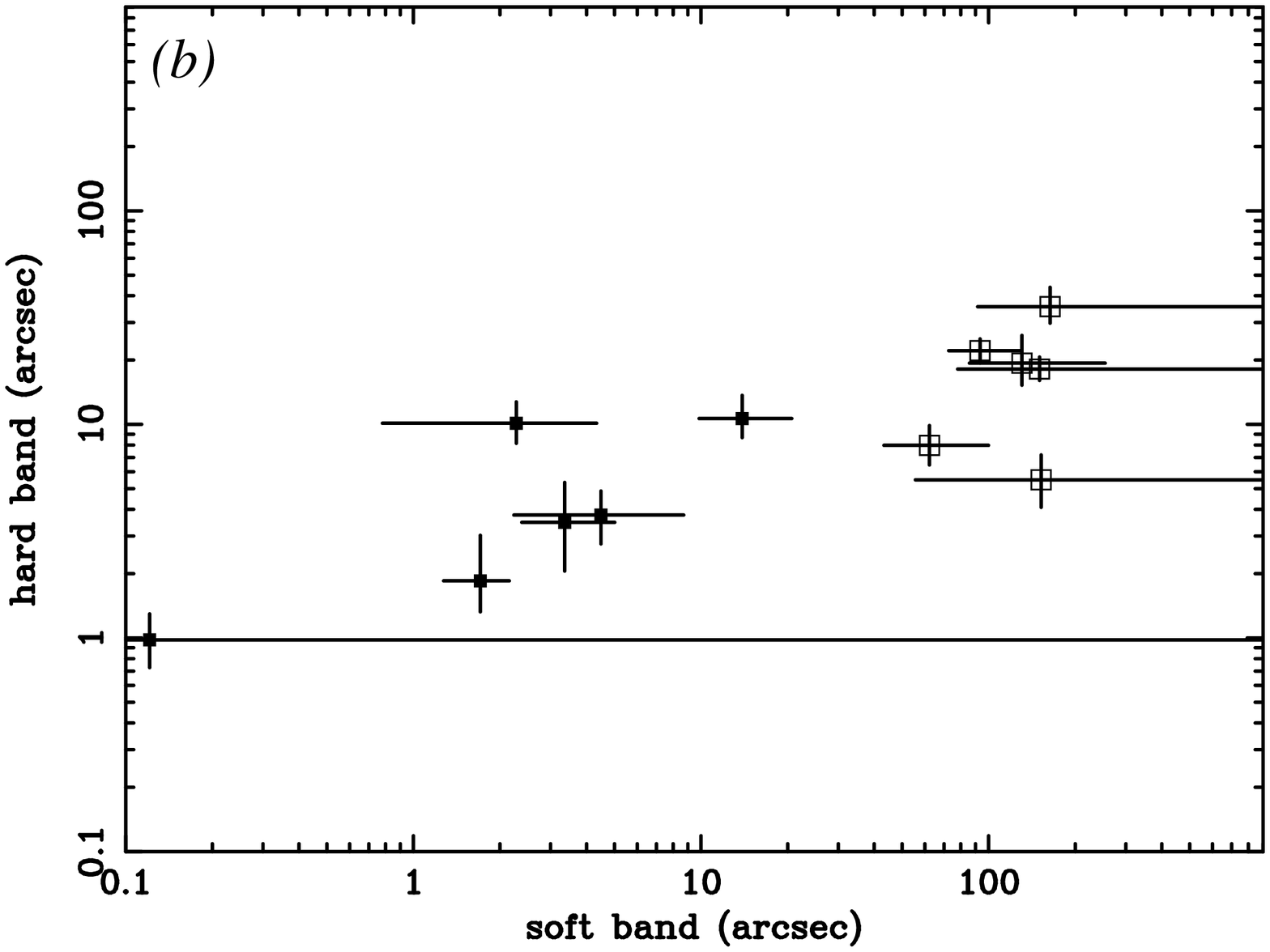}
\caption{(a) The relation between $w_{\rm u}$ and $w_{\rm d}$.
The close and open triangles are for the thermal and non-thermal
emissions, respectively.
(b) The relation between $w^{\rm s}$ and $w^{\rm h}$.
The close and open boxes are in the upstream and downstream, respectively.
\label{relation}}
\end{figure}

Figure~\ref{relation}(a) shows the relation of the scale widths
between the downstream and upstream sides for each filament.
Although there is a large scatter,
$w_{\rm u}$ is systematically smaller than 
$w_{\rm d}$ in both the non-thermal and thermal emissions.

Figure~\ref{relation}(b) shows the relation
between $w^{\rm s}$ and $w^{\rm h}$.
We find that $w_{\rm d}^{\rm s}$ is significantly larger than 
$w_{\rm d}^{\rm h}$,
whereas $w_{\rm u}^{\rm s}$ and $w_{\rm u}^{\rm d}$ are comparable 
with each other. The mean values are  
$\overline{w_{\rm u}^{\rm s}} = 4.3$~arcsec = 0.04~pc, 
$\overline{w_{\rm d}^{\rm s}} = 1.3\times 10^2$~arcsec = 1.1~pc,
$\overline{w_{\rm u}^{\rm h}} = 5.1$~arcsec = 0.04~pc,
and $\overline{w_{\rm d}^{\rm h}} = 18$~arcsec = 0.2~pc.
Note that the minimum value of $w_{\rm u}^{\rm h}$ is
only 0.98~arcsec = 0.01~pc (filament 4).
Maybe since their wide scatter is due to that
these are the projected values of
the possible sheet-like with wavy and/or curved shape.

\section{Discussion}

\subsection{Thermal Plasma}
\label{discuss1}

Although the best-fit NEI model in Table~\ref{thermal_para}
is rejected statistically,
it globally fits the thermal emission of SN~1006 as shown
in Figure~\ref{spectrum}.
The temperature ($kT$=0.24~keV) is similar to the results obtained by
\citet{vink}, \citet{dyer}, and \citet{allen},
but lower than that by \citet{koyama1995}.
Since the spatial resolution of {\it Chandra} enables us
to remove the non-thermal photons
from the thermal emission more accurately than the {\it ASCA} case,
the present results should give a more precise description on
the thermal plasma.
Like the previous observations \citep{koyama1995,allen},
heavy elements, in particular iron, are overabundant,
which implies that the X-ray emitting thermal plasma is dominated
by the ejecta from type Ia SN.
The fact that the thermal emission is enhanced at the inner shell region also 
suggests the ejecta origin.

From the emission measure ($E.M.$) and 
assuming a uniform density plasma of a prorate shape
with the 3-axis radii of 140, 120, and 120~arcsec, 
we estimate the density $n_{\rm e}$ in the inner region to be 0.36~cm$^{-3}$.
Then from the best-fit ionization parameter,
$\tau$ is $2.6\times 10^{10}$~s = $8.3\times 10^2$~yr, 
roughly consistent with the age of SN~1006.

Even in the soft2 band profiles,
the thermal components are not prominent (see Figure~\ref{profiles}),
which prevents us from high quality study for
the morphology of the thermal plasma.
Nevertheless, we found that the profiles of thermal filaments are
largely anti-symmetric. 
The scale width  $w_{\rm u}^{\rm s}$ is very sharp
and comparable to $w_{\rm u}^{\rm h}$,
whereas $w_{\rm d}^{\rm s}$ shows a relatively large scale width.
Although $w_{\rm u}^{\rm s}$ couples to $x_{\rm 0}$,
which is frozen to the best-fit value in the hard band
(see Figure~\ref{relation}(b)),
we can say that the thermal shock has very sharp rise in upstream.

The downstream scale width $w_{\rm d}^{\rm s}$ is comparable to the shock 
width derived from the Sedov solution of about 75~arcsec = 0.7~pc for SN~1006
having a 15~arcmin = 8~pc radius \citep{green}.
Therefore, the thermal filaments may nicely trace 
the density profiles of the Sedov solution.

\subsection{Non-Thermal Filaments} 
\label{discuss2}

In this section, we interpret the scale width of hard band X-rays in upstream,
$w_{\rm u}^{\rm h}$.

The hardest spectral regions, the non-thermal filaments,
are the most probable site of the maximum energy acceleration.
From the observation alone, we can not judge 
whether the filaments are strings or sheets in edge-on configuration.
\citet{hester} suggested that thin sheet-like shock fronts are 
seen as filaments on the edge of the SNR.
The filaments seen in figures of \citet{hester}
resemble the structure in SN~1006.
Furthermore, filaments should be also seen in the inner part of the 
shell if they are strings, however it is not the observed case.
Therefore, we assume that the filaments have sheet-like structure
normal to the shock direction.
The depth of the sheet is unclear
but would be similar to or smaller than the length of the filament.  
We thus assume the depth of the sheet to be about 1~pc
in the following discussion.

The most likely scenario of cosmic ray acceleration at the SNR shock
is diffusive shock acceleration (DSA).
The predicted results from this model however are
highly dependent on many parameters of the magnetic field,
such as the angle between the magnetic field and shock normal direction,
magnetic field strength and its fluctuation.
Here, we investigate the observed profiles based on a DSA model, 
and estimate the physical quantities
such as diffusion coefficient, the direction of magnetic field,
maximum energy of accelerated electrons, and the injection efficiency.
Although a realistic condition may be more complex as is 
discussed by \citet{ellison} and \citet{berezhko},
we at first apply a simplest DSA model that
the back reaction of accelerated particles is neglected.

\subsubsection{Diffusion coefficient in upstream}	
\label{DSA}

The scale widths in the upstream side are largely
scattered in the range of 0.98--11~arcsec
(Table~\ref{pro_para}).
Since this large scatter would be due to the projected 
effect of possible sheet-like structure with wavy and/or curved shape,
real widths should be smaller than the observed (projected) width.
To be conservative, we however adopted the mean value of 4.3~arcsec,  
or 0.04~pc at 1.8~kpc distance in the following discussion.

We use the results of {\it srcut},
the fit of the wide band spectra from X-ray to radio band.
The high energy roll-off $\nu_{\rm rolloff}$ {\it srcut} is
the consequent of either age-limit (acceleration dominant),  
synchrotron loss, or diffusive escape (see \S~\ref{srcut}).
The best-fit $\nu_{\rm rolloff}$ at the filaments is
$2.6_{-0.7}^{+0.7}\times 10^{17}$~Hz,
which can be converted to $1.1_{-0.3}^{+0.3}$~keV.
These are consistent with the {\it ASCA} result of
3.0$_{-0.2}^{+0.1}\times 10^{17}$~Hz or $1.24_{-0.08}^{+0.04}$~keV
by \citet{dyer}.

Since most of the non-thermal X-ray photons are observed
in the downstream region,
the synchrotron radiation is mainly due to the downstream region.
Using the best-fit $\nu_{\rm rolloff}$,
we constrain the maximum energy of electrons $E_{\rm max}$
and magnetic field in downstream $B_{\rm d}$ from the eq.(\ref{rolloff});
\begin{equation}
E_{\rm max}B_{\rm d}^{0.5} = 0.37_{-0.06}^{+0.04}\ \ \  {\rm [ergs\ G^{0.5}]} .
\label{E_peak}
\end{equation}
In the case of the strong shock,
the magnetic field in downstream $B_{\rm d}$ can be related to
that in upstream $B_{\rm u}$ as
\begin{equation}
B_{\rm d} = \left(\cos^2\theta_{\rm u} + r^2\sin^2\theta_{\rm u}\right)^{\frac{1}{2}}B_{\rm u}\label{r},
\end{equation}
where $r$ and $\theta_{\rm u}$ are the compression ratio
and the magnetic field angle to the shock normal direction in upstream.

The diffusion coefficients in upstream ($K_{\rm u}$)
is estimated from eq.(\ref{Ku})
($u_{\rm u} = u_{\rm s}$)
as following;
\begin{eqnarray}
K_{\rm u} &\simeq& w_{\rm u}^{\rm h}\cdot u_{\rm u} = \overline{w_{\rm u}^{\rm h}}\cdot u_{\rm s} \nonumber \\
&=& 3.1\times 10^{25}\ \ \ {\rm [cm^2s^{-1}]}, \label{Ks1}
\end{eqnarray}
where the shock speed $u_{\rm s}$ is assumed to be 2600~km~s$^{-1}$
\citep{laming}.
The diffusion coefficient in upstream is given from 
Using eq.(\ref{K_obl}),
we can derive $K_{\rm u}$ for the electrons of $E_{\rm max}$ as,
\begin{equation}
K_{\rm u} = \frac{1}{3}\xi_{\rm u}\left(\cos^2\theta_{\rm u}+\frac{\sin^2\theta_{\rm u}}{1+\xi_{\rm u}^2}\right)\frac{E_{\rm max}}{{\rm e}B_{\rm u}}c ,
\label{K_u}
\end{equation}
where $\xi_{\rm u}(>1)$ is the fluctuation of the magnetic field.

We can derive following equations
from eq.(\ref{E_peak}), (\ref{r}), (\ref{Ks1}), and (\ref{K_u}),
the maximum energy ($E_{\rm max}$) and magnetic field in upstream
($B_{\rm u}$) are,
\begin{eqnarray}
E_{\rm max} &=& 37\xi_{\rm u}^{-\frac{1}{3}}\left(\cos^2\theta_{\rm u} + \frac{\sin^2\theta_{\rm u}}{1+\xi_{\rm u}^2}\right)^{-\frac{1}{3}}\nonumber \\
&&\times \left(\cos^2\theta_{\rm u}+r^2\sin^2\theta_{\rm u}\right)^{-\frac{1}{6}} {\rm [TeV]},\label{Emax}\\
B_{\rm u} &=& 4.0\times 10^{-5}\xi_{\rm u}^{\frac{2}{3}}
\left(\cos^2\theta_{\rm u} + \frac{\sin^2\theta_{\rm u}}{1+\xi_{\rm u}^2}\right)^\frac{2}{3}
\nonumber \\
&&\times\left(\cos^2\theta_{\rm u}+r^2\sin^2\theta_{\rm u}\right)^{-\frac{1}{6}}
{\rm [G]}.\label{B}
\end{eqnarray}

Since SN~1006 is a type Ia SN located at a high latitude of 460~pc 
above the Galactic plane ($l=14.6$ at 1.8~kpc distance),
the interstellar magnetic field would be smaller than
the typical value in the Galactic plane
of the order of 10~$\mu$G.
We therefore conservatively assume that $B_{\rm u}$ = 10~$\mu$G,
which is slightly larger than the combined results of
the X-ray and TeV Gamma-rays  \citep{tanimori2001}.
Then eq.(\ref{B}) becomes,
\begin{equation}
\xi_{\rm u}^{\frac{2}{3}}\left(\cos^2\theta_{\rm u} + \frac{\sin^2\theta_{\rm u}}{1+\xi_{\rm u}^2}\right)^\frac{2}{3}\left(\cos^2\theta_{\rm u}+r^2\sin^2\theta_{\rm u}\right)^{-\frac{1}{6}} = 0.25.
\label{xi_B}
\end{equation}

For the parallel magnetic field ($\theta_{\rm u} = 0^\circ$),
$\xi_{\rm u}$ becomes smaller than 1.
This unrealistically small value of $\xi_{\rm u}$ conflicts
to a usual DSA which assume a parallel magnetic field configuration.

An alternative scenario is oblique magnetic field to the shock normal
in this region ($\theta_{\rm u} > 0^\circ$).
For simplicity, we neglect the back reaction of accelerated particle,
then the compression ratio becomes 4 for strong shock limit.
The allowed region of $\theta_{\rm u}$ is given from the eq.(\ref{xi_B}) as;
\begin{equation}
\theta_{\rm u} \geq 82^\circ .
\label{theta_r=4}
\end{equation}
Thus the magnetic field in upstream near the filaments is
almost perpendicular.

Assuming  $\theta_{\rm u} = 90^\circ$, $r = 4$, and $B_{\rm u}$~=~10~$\mu$G,
$E_{\rm max}$ is given from eq.(\ref{Emax}),   
\begin{eqnarray}
E_{\rm max} &=& r^{-\frac{1}{2}}B_{\rm u}^{-\frac{1}{2}}\cdot 0.23_{-0.04}^{+0.03}\ \ {\rm [TeV]} \nonumber \\
&=& 37_{-7}^{+4}\ \ {\rm [TeV]}.
\end{eqnarray}

When particles are accelerated very efficiently,
their back reactions to the shock can not be ignored
hence $r$ becomes larger than 4.
Even if the compression ratio is the largest of
$r=7$ \citep{ellison,berezhko},
the allowed range is  $\theta_{\rm u} > 80^\circ$.
Therefore, we can safely predict that
the magnetic field in the NE shell of SN~1006 is
nearly perpendicular to the shock normal.

As for the downstream region,
the observed spatial profile seems to be incompatible
with the solution derived by \citet{blandford1978}.
The maximum electron energy $E_{\max}$ would be determined by
the balance of the time scales
between the accelerating and the synchrotron cooling.
These time scales may depend on the structure and the fluctuation
of the magnetic field along the shock normal;
both time scales become smaller
with larger magnetic field, while the former becomes also 
smaller with larger fluctuation of the magnetic field.

The shock flow may compresses 
and partly stretches the magnetic field in the radial direction,
which may produce highly disordered magnetic field
with small fraction of radial component,
as discussed by \citet{reynolds1993} with the radio polarization data,
who reported that only 15--20\% of the magnetic-field energy
in SN~1006 NE shell in radial polarization,
and most of the magnetic field is disordered
with the scale smaller than 0.2~pc.

To determine the magnetic field in downstream,
we thus need require more complicated processes
such as the history of the shock propagation and the non-linear effect,
and many assumptions,
which is beyond this paper and would leave for a future study.

\subsubsection{Injection Efficiency}
\label{discuss3}

We define the injection efficiency
$\eta \equiv \frac{n_{\rm e}^{\rm NT}}{n_{\rm e}^{\rm T}}$, 
where $n_{\rm e}^{\rm NT}$ and $n_{\rm e}^{\rm T}$ are 
the number densities of non-thermal and thermal electrons in the filaments.
The depth of the filament (sheet-like) is assumed 
to be 1~pc (see \S~\ref{DSA}), with uniform electron density.
 
The non-thermal electron flux (energy) are estimated
using the method given in Appendix (\S~\ref{srcut}), 
where the non-thermal X-ray flux and spectra of each filament
are taken from Table~\ref{pro_para} and Table~\ref{spec_para}.
We adopt the minimum energy of non-thermal electrons (injection energy)
$E_{\rm min}$ to be 0.24~keV
(the temperature of the thermal plasma; see Table~\ref{thermal_para}).
The magnetic field and the maximum energy are unknown parameters,
however we can suggest that
the magnetic field in downstream is significantly larger than that in upstream.
In this section,
we adopt the magnetic field $B$ to be perpendicular to the shock normal,
$B_{\rm d} = 4B_{\rm u}$ = 40~$\mu$G arbitrary
and the maximum energy $E_{\rm  max}$ to be 37~TeV
from the discussion in \S~\ref{DSA}.
The derived number density ($n_{\rm e}^{\rm NT}$),
the total number ($N_{\rm e}^{\rm NT}$),
and the total energy ($E_{\rm e}^{\rm NT}$)
of non-thermal electrons in each filament are
summarized in Table~\ref{injection_para}.

For the estimation of thermal electron flux (energy),
we adopted the projected profile of the thermal X-rays (the soft2 band)
given in Table~\ref{pro_para} and the spectral parameters of thermal 
plasma given in Table~\ref{spec_para}.
The resultant number density ($n_{\rm e}^{\rm T}$),
total number ($N_{\rm e}^{\rm T}$), and total energy ($E_{\rm e}^{\rm T}$) 
of thermal electrons in each filament
are given in Table~\ref{injection_para}.
The injection efficiency 
$\eta\ (\equiv \frac{n_{\rm e}^{\rm NT}}{n_{\rm e}^{\rm T}})$
is then obtained as is shown in Table~\ref{injection_para}.
All the values $\eta$ are nearly identical in each filament of
$\sim 1 \times 10^{-3}$
but are about 2 times larger than $5\times 10^{-4}$
derived from the {\it ASCA} data by \citet{allen},
although they also assumed as $B_{\rm d} = 40~\mu$G arbitrary.  
For comparison, we estimate $\eta$ with larger magnetic field of
$B_{\rm d}= 40~\mu$G, and find a larger $\eta$
than \citet{allen}.
They estimated the number density of non-thermal electrons
from larger regions than the filaments,
because {\it ASCA} could not resolve the filaments.
On the other hand, we found that the non-thermal electrons are confined
in the thin filaments,
which are the sites of ongoing acceleration of the non-thermal electrons.
We thus regard that
the present {\it Chandra} result suggests that
the injection occurs more locally and more efficiently
and as a result,
it must be more realistic estimation of $\eta$ than that by \citet{allen}.  

The energy densities of the magnetic field,
the thermal plasma, and the non-thermal electrons in the filaments
are  $6.4\times 10^{-11}$~ergs~cm$^{-3}$,
$2.6\times 10^{-10}$~ergs~cm$^{-3}$,
and $6.9\times 10^{-11}$~ergs~cm$^{-3}$, respectively.
Thus, at the shocked region,
the magnetic field and non-thermal electrons
are in energy equi-partition
and are slightly smaller than the thermal energy (about 30\%).

As suggested by \citet{ellison}, 
the non-thermal protons should carry larger energy than 
electrons, hence particle energy becomes larger than that of magnetic
and possibly that of thermal.
We therefore must consider non-linear effects 
suggested by \citet{ellison} and by \citet{berezhko},
however more quantitative scenario including the non-linear
effects must be a future work.
  
\section{Summary}

(1) X-ray emissions from the NE shell of SN~1006 are found to be 
composites of filaments, clumps and more extended diffuse emissions. 
We resolved the thermal and non-thermal X-rays using the different
spectral shape and morphology.
 
(2) The spectrum of the thermal component has $kT$ = 0.24~keV temperature
and relatively small ionization time scale
$n_{\rm e}\tau = 1.1\times 10^{10}$~s~cm$^{-3}$.
The chemical compositions are overabundant, especially in iron, which
suggests that the X-ray emitting plasma originates from ejecta
and the progenitor is type~Ia.

(3) The non-thermal components can be described with a power-law function
with the photon index 2.1--2.3 in the narrow filaments
and 2.5 at the inner region of the shell.

(4) The structure of the filaments shows different characteristics
in thermal and non-thermal X-rays.
The thermal plasma has the scale width of 1~pc in downstream,
similar to the shock width derived from Sedov equations. 
The non-thermal filaments show extremely small scale width
of $w_{\rm u}^{\rm h} \sim$~0.04~pc and $w_{\rm d}^{\rm h} \sim$~0.2~pc,
in upstream and downstream, respectively.

(5) In a diffusive shock acceleration model,
the observed thin filaments requires a nearly perpendicular magnetic field
with the angle between the magnetic field in upstream 
(assumed as 10~$\mu$G) and the shock normal
of larger than $\sim 80^\circ$.
The maximum energy of the electrons are 30--40~TeV.

(6) The injection efficiency is estimated to be $\eta$ = 1$\times 10^{-3}$, 
suggesting that thermal particles are injected locally and very effectively.
Then the energy density of non-thermal electrons becomes comparable to
that of the magnetic field and about 30\% of the thermal energy density.
Thus non-linear effect of the shock structure
and acceleration mechanism must be considered.

\acknowledgements

We thank the anonymous referee for his/her helpful comments.
Our particular thanks are due to
M. Hoshino, T. Terasawa, T. Yoshida, and S. Inutsuka
for their fruitful discussions and comments.
A.B. and M.U. are supported by JSPS Research Fellowship for Young Scientists.

\appendix

\section{Appendix}
\label{appendix}

In this appendix,
we briefly introduce the relevant software tools and equations,
which are used for the discussions in the text.  

\subsection{Electron Spectra and {\it srcut} Model}
\label{srcut}

The spectrum of non-thermal electrons accelerated
by the diffusive shock is \citep{bell};
\begin{eqnarray}
\frac{{\rm d}n_{\rm e}^{\rm NT}}{{\rm d}E^{\rm NT}} &=&
\kappa (E^{\rm NT}+m_{\rm e}c^2)({E^{\rm NT}}^2
+2m_{\rm e}c^2E^{\rm NT})^{-\frac{(p+1)}{2}}
\times \exp(-E^{\rm NT}/E_{\rm max})
\label{NT_spec},
\end{eqnarray}
where $n_{\rm e}^{\rm NT}$ and $E^{\rm NT}$
are the number density and the energy 
of non-thermal electrons, respectively.

The synchrotron radiation power per unit frequency from a single electron of 
energy $E$ in a magnetic field $B$
is
\begin{equation}
P(\nu,\alpha) = \frac{\sqrt{3}{\rm e}^3B\sin\alpha}{m_{\rm e}c^2}
F\left(\frac{\nu}{\nu_c}\right),
\label{power}
\end{equation}
where $\alpha$ and  $F(x)$ are
the pitch angle and the function given by \citet{rybicki}.
The peak frequency $\nu_c$ is
\begin{equation}
\nu_c = \frac{3cE^2{\rm e}B\sin\alpha}{4\pi(m_{\rm e}c^2)^3}.
\end{equation}

Convolving the eq.(\ref{NT_spec}) with (\ref{power}),
we obtain the spectrum of synchrotron radiation in the pitch 
angle $\alpha$ as,
\begin{equation}
f(\nu,\alpha) = \int_{E_{\rm min}}^\infty{\rm d}E^{\rm NT}\frac{{\rm d}n_{\rm e}^{\rm NT}}{{\rm d}E_{\rm e}^{\rm NT}}P(\nu,\alpha).
\end{equation}
Averaged over the pitch angle, 
we finally obtain the synchrotron radiation energy per unit volume,
frequency, and time,
\begin{equation}
f(\nu) = \frac{1}{2}\int_0^\pi{\rm d}\alpha\sin\alpha f(\nu,\alpha) .
\end{equation}

The observed spectrum is fitted with this model spectrum using the 
{\it Chandra} software 
{\it srcut} with 3 free parameters \citep{reynolds1998,reynolds1999}.
The normalization constant and spectral index are determined
so as to reproduce the  flux and slope in the radio band.
These parameters are converted to $\kappa$ and $p$
in eq.(\ref{NT_spec}).
The other fitting parameter $\nu_{\rm rolloff}$ is
a function of $E_{\rm max}$ in eq.(\ref{NT_spec})
and magnetic field ($B$) as is shown below  \citep{reynolds1999};

\begin{equation}
\nu_{\rm rolloff} = 0.5\times10^{16}\left(\frac{B}{{\rm 10~\mu G}}\right)
\left(\frac{E_{\rm max}}{{\rm 10~TeV}}\right)^2\ \ \ {\rm [Hz]} .
\label{rolloff}
\end{equation}
This equation gives constraint on the maximum electron energy $E_{\rm max}$
and magnetic field $B$.

Since available radio data of SN~1006 have spatial resolution far larger
than the scale of the X-ray filaments, we have no accurate radio flux
nor index from the position of the X-ray filaments. We therefore  fixed 
the radio index to the poor resolution radio result of $p=2.14$
by \citet{allen} 
and the radio flux is treated as a free parameter
for the present {\it srcut} fitting.
Allowing the radio index to vary from  $p=2.0$ to $p=2.2$,
we see no significant difference in the best-fit parameters
within the statistical errors.

\subsection{Diffusion versus Advection}
 
The spatial structure of the relativistic electrons
produced by the diffusive shock acceleration across the shock is determined
by the competing process of diffusion and advection. 
The advection time scale ($\tau_{\rm ad}$) is given by
\begin{equation}
\tau_{\rm ad} = \frac{w}{u},
\end{equation}
where $w$ is the scale width of the spatial distribution of
the relativistic electrons and $u$ is the flow speed.  

The diffusion time scale ($\tau_{\rm dif}$) is given from
the random-walk theory as;
\begin{equation}
\tau_{\rm dif} = \frac{w^2}{K},
\label{t_diff}
\end{equation}
where $K$ is the diffusion coefficient.

In order that the high energy electrons are accelerated at shock front,
electrons in upstream should be diffused back to the shock front 
against the advection to the downstream side (shock flow).
Therefore, $\tau_{\rm ad}$ should be nearly equal to $\tau_{\rm dif}$, 
hence  $\frac{w_{\rm u}}{u_{\rm u}} \simeq \frac{w_{\rm u}^2}{K_{\rm u}}$.
We thus obtain;  
\begin{equation}
w_{\rm u} \simeq \frac{K_{\rm u}}{u_{\rm u}}.
\label{Ku}
\end{equation}
For more exact formalisms, see e.g. \citet{bell}, \citet{blandford1978}, and
\citet{drury1983}.  

Let the mean free path parallel to the magnetic field be
a constant factor $\xi$ times the gyro radius
$r_{\rm g}$.
Then, the effective diffusion coefficient along the shock normal is given as
\citep{jokipii,skilling}
\begin{eqnarray}
K &=& \frac{1}{3}\xi r_{\rm g}c\left(\cos^2\theta_{\rm u} + \frac{\sin^2\theta_{\rm u}}{1+\xi^2}\right),
\label{K_obl}\\
r_{\rm g} &=& \frac{E_{\rm e}} {{\rm e}B},
\end{eqnarray}
where $e$, $E_{\rm e}$, $B$, and $\theta_{\rm u}$ are the electric charge, 
the energy of relativistic electrons, and the magnetic field, 
and the angle between the magnetic field in upstream
and the shock normal, respectively.

\onecolumn

\begin{deluxetable}{p{9pc}c}
\tabletypesize{\scriptsize}
\tablecaption{The best-fit parameters of the spectrum of the inner region\tablenotemark{a}.
\label{thermal_para}}
\tablewidth{0pt}
\tablecolumns{1}
\tablehead{
\colhead{Parameters} & \colhead{Best-fit value}}
\startdata
Power-law Model\\
\hspace*{5mm}Photon Index\dotfill & 2.51 (2.48--2.53) \\
\hspace*{5mm}Flux\tablenotemark{b}\ [$\rm ergs\ cm^{-2}\ s^{-1}$]\dotfill & $3.8 \times 10^{-12}$ \\
NEI model\\
\hspace*{5mm}Temperature $[{\rm keV}]$\dotfill & 0.24 (0.21--0.26) \\
\hspace*{5mm}Abundances\tablenotemark{c} & \\
\hspace*{10mm}C \dotfill & ($<$0.1) \\
\hspace*{10mm}N \dotfill & ($<$0.03) \\
\hspace*{10mm}O \dotfill & 3.3 (3.0--3.5) \\
\hspace*{10mm}Ne \dotfill & 4.8 (4.4--5.2) \\
\hspace*{10mm}Mg \dotfill & 51 (42--61) \\
\hspace*{10mm}Si \dotfill & 131 (121--140) \\
\hspace*{10mm}S \dotfill & 10 (6.8-13) \\
\hspace*{10mm}Fe \dotfill & 37 (28--46) \\
\hspace*{5mm}$n\tau$ $\rm [\times 10^{9}\ s\ cm^{-3}]$\dotfill & 10.8 (9.9--11.1) \\
\hspace*{5mm}$E.M.$\tablenotemark{d} [$\times 10^{55}$cm$^{-3}$]\dotfill & 2.1 (1.9--2.2) \\
\hspace*{5mm}Flux\tablenotemark{b}\ [$\rm ergs\ cm^{-2}\ s^{-1}$]\dotfill & 3.3$\times 10^{-12}$ \\
$N_{\rm H}$ [$\rm 10^{20}\ H\ cm^{-2}$]\dotfill & 9.0 (8.5--9.4) \\
Reduced $\chi^2$ [$\chi^2$/d.o.f]\dotfill & 447.9/215 \\
\enddata
\tablenotetext{a}{Parentheses indicate single-parameter
90\% confidence regions.}
\tablenotetext{b}{In the 0.3--10.0~keV band.}
\tablenotetext{c}{Abundance ratio relative to the solar value \citep{anders}.}
\tablenotetext{d}{$E.M. = n_{\rm e}^2V$,
where $n_{\rm e}$ and $V$ are the electron density and the volume,
respectively.} 
\end{deluxetable}

\begin{deluxetable}{p{1.5pc}cccccccc}
\tabletypesize{\scriptsize}
\tablecaption{The best-fit parameters of the profiles of the filaments.%
\tablenotemark{a}
\label{pro_para}}
\tablewidth{0pt}
\tablecolumns{2}
\tablehead{
\colhead{No.} & \colhead{$A^{\rm h}$} & \colhead{$w_{\rm u}^{\rm h}$}
 & \colhead{$w_{\rm d}^{\rm h}$} & \colhead{Reduced $\chi^2$} 
& \colhead{$A^{\rm s}$} & \colhead{$w_{\rm u}^{\rm s}$} 
& \colhead{$w_{\rm d}^{\rm s}$} & \colhead{Reduced $\chi^2$}\\
 & [cnts arcsec$^{-1}$]\tablenotemark{b} & [arcsec] &  [arcsec] & [$\chi^2$/d.o.f.] & [cnts arcsec$^{-1}$]\tablenotemark{c} & [arcsec] &  [arcsec] & [$\chi^2$/d.o.f.]
}
\startdata
1\dotfill & 53 & 1.9 & 36 & 119.1/105 & 55 & 1.7 & $1.6\times 10^2$ & 104.0/94 \\
 & (49--58) & (1.3--3.0) & (30--44) & \nodata & (49--61) & (1.3--2.2) & ($>$92) & \nodata \\
2\dotfill & 63 & 11 & 22 & 347.3/296 & 1.0$\times 10^2$ & 14 & 93 & 481.8/298 \\
 & (58--68) & (8.7--14) & (19--25) & \nodata & (95--1.2$\times 10^2$) & (9.8--21) & (73--$1.3\times 10^2$) & \nodata \\
3\dotfill & 66 & 10 & 5.5 & 141.3/114 & 44 & 2.3 & $1.5\times 10^2$ & 119.8/95 \\
 & (58--74) & (8.2--13) & (4.1--7.2) & \nodata & (37--52) & (0.78--4.3) & ($>$56) & \nodata \\
4\dotfill & 68 & 0.98 & 18 & 103.2/99 & 38 & 0.12 & $1.5\times 10^2$ & 106.1/81\\
 & (62--74) & (0.73--1.3) & (16--21) & \nodata & (34--42) & (not determined) & ($>$78) & \nodata \\
5\dotfill & 75 & 3.8 & 8.0 & 218.3/166 & 33 & 4.5 & 62 & 236.5/167\\
 & (66--85) & (2.8--4.9) & (6.5--9.9) & \nodata & (24--41) & (2.2--8.7) & (43--$1.0\times 10^2$) & \nodata \\
6\dotfill & 87 & 3.5 & 19 & 99.9/86 & 1.1$\times 10^2$ & 3.4 & $1.3\times 10^2$ & 118.4/87 \\
 & (79--96) & (2.1--5.4) & (15--26) & \nodata & (1.0$\times 10^2$--1.3$\times 10^2$) & (2.4--5.0) & (86--$2.5\times 10^2$) & \nodata \\
\enddata
\tablenotetext{a}{Parentheses indicate single-parameter
90\% confidence regions.}
\tablenotetext{b}{In the 2.0--10.0~keV band.}
\tablenotetext{c}{In the 0.4--0.8~keV band.}
\end{deluxetable}

\begin{deluxetable}{p{8pc}ccccccc}
\tabletypesize{\scriptsize}
\tablecaption{The best-fit parameters of the spectral fittings
for the filaments.\tablenotemark{a}
\label{spec_para}}
\tablewidth{0pt}
\tablecolumns{2}
\tablehead{
\colhead{Parameters} & \colhead{1} & \colhead{2} & \colhead{3} & \colhead{4} & \colhead{5} & \colhead{6} & \colhead{Total}
}
\startdata
Power-low model\\
\hspace*{3mm}$\Gamma$\dotfill & 2.0 (1.8--2.2) & 2.4 (2.3--2.6) & 2.3 (2.1--2.4) & 2.3 (2.1--2.4) & 2.3 (2.1--2.5) & 2.3 (2.1--2.4) & 2.31 (2.29--2.33)\\
\hspace*{3mm}$N_{\rm H}$ [$\rm 10^{21}\ H\ cm^{-2}$]\dotfill & 1.5 (0.9--2.1) & 1.8 (1.5--2.0) & 1.2 (1.0--1.5) & 1.6 (1.3--2.0) & 1.4 (1.1--1.8) & 1.4 (1.1--1.7) &1.6 (1.6--1.7) \\
\hspace*{3mm}Flux\tablenotemark{b}  [$\rm ergs\ cm^{-2}\ s^{-1}$]\dotfill & $3.7\times 10^{-13}$ & $5.4\times 10^{-13}$ & $3.7\times 10^{-13}$ & $2.5\times 10^{-13}$ & $2.3\times 10^{-13}$ & $4.2\times 10^{-13}$ & 1.8$\times 10^{-12}$  \\
\hspace*{3mm}Reduced $\chi^2$ [$\chi^2$/d.o.f]\dotfill & 44.1/43 & 232.5/164 & 196.3/140 & 139.8/125 & 143.8/125 & 173.4/130 & 273.3/211 \\
{\it srcut} model\\
\hspace*{3mm}$\nu_{\rm rolloff}$ [$\times 10^{17}$Hz]\dotfill & 13 (3.6--94) & 1.7 (1.1--2.3) & 2.8 (1.8--4.7) & 2.9 (1.5--6.2) & 2.4 (1.3--5.0) & 2.8 (1.7--5.0) & 2.6 (1.9--3.3) \\
\hspace*{3mm}$N_{\rm H}$ [$\rm 10^{21}\ H\ cm^{-2}$]\dotfill & 1.3 (1.0--1.7) & 1.4 (1.2--1.5) & 1.0 (0.8--1.1) & 1.3 (1.1--1.5) & 1.1 (0.9--1.4) & 1.1 (0.9--1.3) & 1.3 (1.2--1.5) \\
\hspace*{3mm}Flux\tablenotemark{b}  [$\rm ergs\ cm^{-2}\ s^{-1}$]\dotfill & $3.6\times 10^{-13}$ & $5.3\times 10^{-13}$ & $3.6\times 10^{-13}$ & $2.4\times 10^{-13}$ & $2.2\times 10^{-13}$ & $4.1\times 10^{-13}$ & $1.8\times 10^{-12}$ \\
\hspace*{3mm}Reduced $\chi^2$ [$\chi^2$/d.o.f]\dotfill & 43.7/43 & 225.1/164 & 190.9/140 & 139.1/125 & 141.1/125 & 170.7/130 & 265.2/211\\
\enddata
\tablenotetext{a}{Parentheses indicate
single-parameter 90\% confidence regions.}
\tablenotetext{b}{In the 0.3--10.0~keV band.}
\end{deluxetable}

\begin{deluxetable}{p{9pc}ccccccc}
\tabletypesize{\scriptsize}
\tablecaption{The number densities and energies of thermal and non-thermal electrons
\label{injection_para}} 
\tablewidth{0pt}
\tablecolumns{2}
\tablehead{
\colhead{Filament No.} & \colhead{1} & \colhead{2} & \colhead{3} & \colhead{4} & \colhead{5} & \colhead{6} & \colhead{Total}
}
\startdata
Thermal electrons\\
\hspace*{5mm}$E.M.$ [cm$^{-3}$]\dotfill & $7.0\times 10^{53}$ & $1.0\times 10^{54}$ & $1.3\times 10^{53}$ & $2.5\times 10^{53}$ & $1.3\times 10^{53}$ & $8.4\times 10^{53}$ & $3.1\times 10^{54}$ \\
\hspace*{5mm}Density ($n_{\rm e}^{\rm T}$)\tablenotemark{a} [cm$^{-3}$]\dotfill & 0.43 & 0.51 & 0.26 & 0.41 & 0.30 & 0.60 & 0.45 \\
\hspace*{5mm}Total number ($N_{\rm e}^{T}$)\dotfill & $1.6\times 10^{54}$ & $2.1\times 10^{54}$ & $5.0\times 10^{53}$ & $6.1\times 10^{53}$ & $4.3\times 10^{53}$ & $1.4\times 10^{54}$ & $6.8\times 10^{54}$ \\
\hspace*{5mm}Energy ($E_{\rm e}^{\rm T}$)\tablenotemark{b} [ergs]\dotfill & $9.2\times 10^{44}$ & $1.2\times 10^{45}$ & $2.8\times 10^{44}$ & $3.5\times 10^{44}$ & $2.4\times 10^{44}$ & $7.9\times 10^{44}$ & $3.9\times 10^{45}$\\
Non-thermal electrons\tablenotemark{c}\\
\hspace*{5mm}Density ($n_{\rm e}^{\rm NT}$)\tablenotemark{a} [cm$^{-3}$]\dotfill & $4.7\times 10^{-4}$ & $5.5\times 10^{-4}$ & $7.9\times 10^{-4}$ & $7.1\times 10^{-4}$ & $6.5\times 10^{-4}$ & $7.6\times 10^{-4}$ & $6.2\times 10^{-4}$ \\
\hspace*{5mm}Total number ($N_{\rm e}^{\rm NT}$)\dotfill & $1.8\times 10^{51}$ & $2.2\times 10^{51}$ & $1.5\times 10^{51}$ & $1.1\times 10^{51}$ & $9.4\times 10^{50}$ & $1.8\times 10^{51}$ & $9.3\times 10^{51}$ \\
\hspace*{5mm}Energy ($E_{\rm e}^{\rm NT}$) [ergs]\dotfill & $2.0\times 10^{44}$ & $2.4\times 10^{44}$ & $1.7\times 10^{44}$ & $1.2\times 10^{44}$ & $1.1\times 10^{44}$ & $2.0\times 10^{44}$ & $1.0\times 10^{45}$ \\
Injection efficiency ($\eta$\tablenotemark{d})\dotfill & 1.1$\times 10^{-3}$ & $1.1\times 10^{-3}$ & $3.0\times 10^{-3}$ & $1.7\times 10^{-3}$ & $2.2\times 10^{-3}$ & $1.3\times 10^{-3}$ & $1.4\times 10^{-3}$
\enddata
\tablenotetext{a}
{We assumed that the depth of the emitting volume is 1~pc
and that the filling factor = 1.}
\tablenotetext{b}
{$E_{\rm e}^{\rm T} = \frac{3}{2}n_{\rm e}V^{\rm NT}kT$.}
\tablenotetext{c}
{Integration from $E_{\rm min}$ = 0.24~keV
to $E_{\rm max}$ = 37~TeV (see text).}
\tablenotetext{d}
{$\eta \equiv \frac{n_{\rm e}^{\rm NT}}{n_{\rm e}^{\rm T}}$ (see text).}
\end{deluxetable}

\end{document}